\documentclass[final,3p,12pt]{elsarticle}

\pdfoutput=1

\usepackage{graphicx}

\usepackage{amssymb}

\journal{Carbon}

\begin{document}

\begin{frontmatter}



\title{Molecular dynamics of cleavage and flake formation during the interaction of a graphite surface with a rigid nanoasperity}


\author{A.V.~Khomenko\corref{cor1}}
\ead{khom@phe.sumdu.edu.ua}
\ead[url]{http://personal.sumdu.edu.ua/khomenko/eng/}

\author{N.V.~Prodanov}
\ead{prodk@rambler.ru}

\cortext[cor1]{Corresponding author: Fax: +38 0542 334058 }

\address{Sumy State University, 2 Rimskiy-Korsakov Str., 40007 Sumy, Ukraine}

\begin{abstract}
Computer experiments concerning interactions between a graphite surface and the rigid pyramidal nanoasperity of a friction force microscope tip when it is brought close to and retracted from the graphitic sample are presented. Covalent atomic bonds in graphene layers are described using a Brenner potential and tip-carbon forces are derived from the Lennard-Jones potential. For interlayer interactions a registry-dependent potential with local normals is used. The behavior of the system is investigated under conditions of different
magnitudes of tip-sample interaction and indentation rates. Strong forces between the nanoasperity and carbon atoms facilitate the cleavage of the graphite surface. Exfoliation, i. e. total removal of the upper graphitic
layer, is observed when a highly adhesive tip is moved relative to the surface at low rates, while high rates cause the formation of a small flake attached to the tip. The results obtained may be valuable for enhancing our understanding of the superlubricity of graphite.

\end{abstract}





\end{frontmatter}


\section{Introduction}
\label{intr}

Graphite, especially in highly oriented pyrolytic form (HOPG), has a special place in nanotribology. It provides atomically flat surfaces that are relatively easy to obtain and has been extensively used in experiments exploring atomic-scale friction and wear. Graphite was one of the first materials probed with the friction force microscope (FFM)~\cite{Mate1995}. Recent experimental studies involving graphitic surfaces include, for example, investigation of motion of carbon nanotubes (CNT)~\cite{Falv2000}, frictional resistance of antimony nanoparticles~\cite{Diet2008} and friction at atomic-scale surface steps~\cite{Hols2008}. Obtained results reveal some interesting phenomena, e.g., the dramatic influence of commensurability between CNT and HOPG surface on friction~\cite{Falv2000}, the existence of frictional duality for nanoparticles with two possible scenarios of sliding one of which is frictionless~\cite{Diet2008} and a direction-dependence of friction at atomic-scale steps~\cite{Hols2008}.

A remarkable feature of a typical friction loop obtained with FFM tungsten tip probing a graphite surface is its atomic periodicity~\cite{Mate1995, Mate2002}, which indicates that sliding process is not uniform, but instead an atomic stick-slip takes place. This is rather surprising because many tip atoms contact the graphite surface and the periodic behavior should have been washed out. To explain the atomic periodicity, the inventor of FFM Mate in the earlier studies suggested that the FFM tip drags a graphite flake across the surface~\cite{Mate1995}. However, in the more recent work~\cite{Mate2002} he states that this explanation fell out of favor once researchers started observing atomically periodic friction on nonlayered materials where flake formation is impossible.

Dienwiebel et al.~\cite{Dien2004} studied atomic-scale friction of tungsten tip sliding over a graphite surface while testing a novel FFM. Besides atomic periodicity of friction loops, they observed the so-called superlubricity, which is manifested in a reduction of friction by orders of magnitude. Average friction force in the mentioned system exhibits strong dependence on the rotation angle of the graphite sample around an axis normal to the sample surface. These dependencies consist of two narrow angular regions with high friction, separated by a wide angular interval with nearly zero friction. The distance between the two friction peaks corresponds well with the $60^{\circ}$ symmetry of individual atomic layers in the graphite lattice. This fact and a good fit of the experimental results to numerical simulations carried out on the modified Tomlinson model~\cite{Dien2004B} were the main reasons to argue that the superlubricity took place between the graphite substrate and a graphite flake, attached to the tip. At the two orientations corresponding to the friction peaks, the flake and substrate lattices
were perfectly aligned, while they were incommensurate for the intermediate angles.

However, in the experiments pertaining to superlubricity, there is no firm evidence for the existence of the flake attached to the tip. The imaging of the FFM tip using high-resolution transmission electron microscopy (HRTEM) did not allow its thorough inspection~\cite{Dien2004,DienThes}. This is due to the ambient conditions of measurements which resulted in coverage of the tip by the amorphous oxide layer. It is almost completely removed by the electron irradiation after several minutes of HRTEM work and the flake might be eliminated with this layer.

The mentioned ambiguity in treatment of the experimental data indicates the necessity of thorough theoretical investigations of cleavage and wear of graphite at the nanoscale. However, most of the existing theoretical models of superlow friction of graphite are, to our knowledge, based on the assumption of the presence of the cleaved graphitic layers~\cite{Dien2004B,DienThes,Mats2005,Fili2008} and there is no theoretical confirmation of the existence of the flake.

The facts listed above provide the impetus for the development of new theoretical models which would be able to clearly suggest the right interpretation of experimental results or even to reproduce the experiments. As the first step towards accomplishing these tasks large-scale classical molecular dynamics (MD) simulations described in this work have been performed. Classical MD is a widely used tool for investigation of friction, wear, and related processes at the atomic scale and it provides insights into these phenomena that could
not have been obtained in any other way~\cite{Khome2008,Khome2009,Harri1991,Landm1989,Landm1990,Harri1992,Dedko2000,Schal2004,HeoSJ2005}. In the current work the interactions of a graphite surface with adhesive absolutely rigid nanoasperity of the FFM tip when it is approached to and retracted from the graphitic substrate are studied under conditions of different magnitudes of tip-sample interaction and indentation rates. The main aim of the present study is to show that MD simulations using realistic empirical potentials are able to reproduce the formation of the flake under appropriate conditions and to shed some light on the accompanying physical processes. The next section describes the details of the simulations.

\section{Simulation setup}
\label{model}

\begin{figure}[htb]
\centerline{\includegraphics[width=0.48\textwidth]{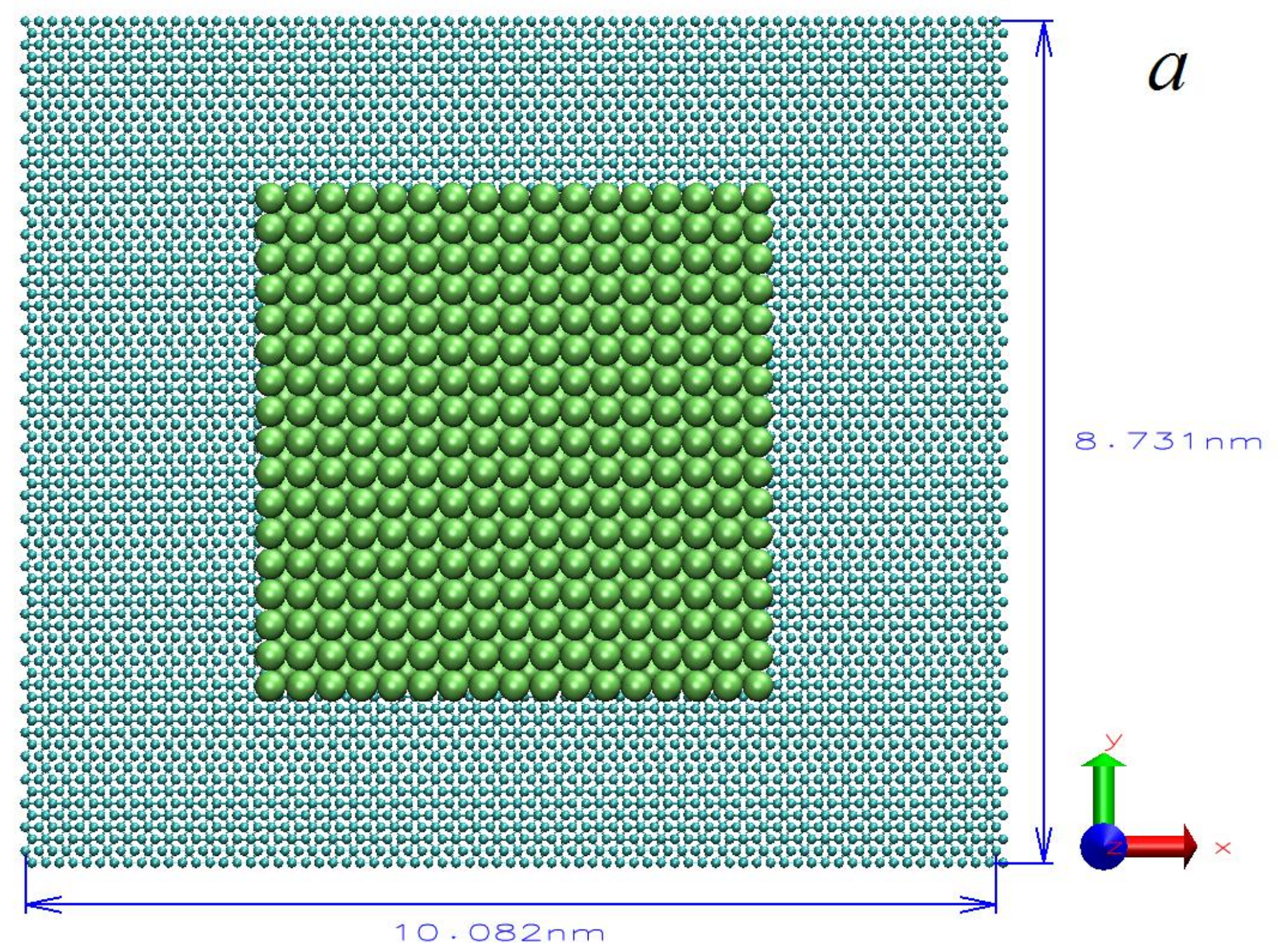}
\includegraphics[width=0.52\textwidth]{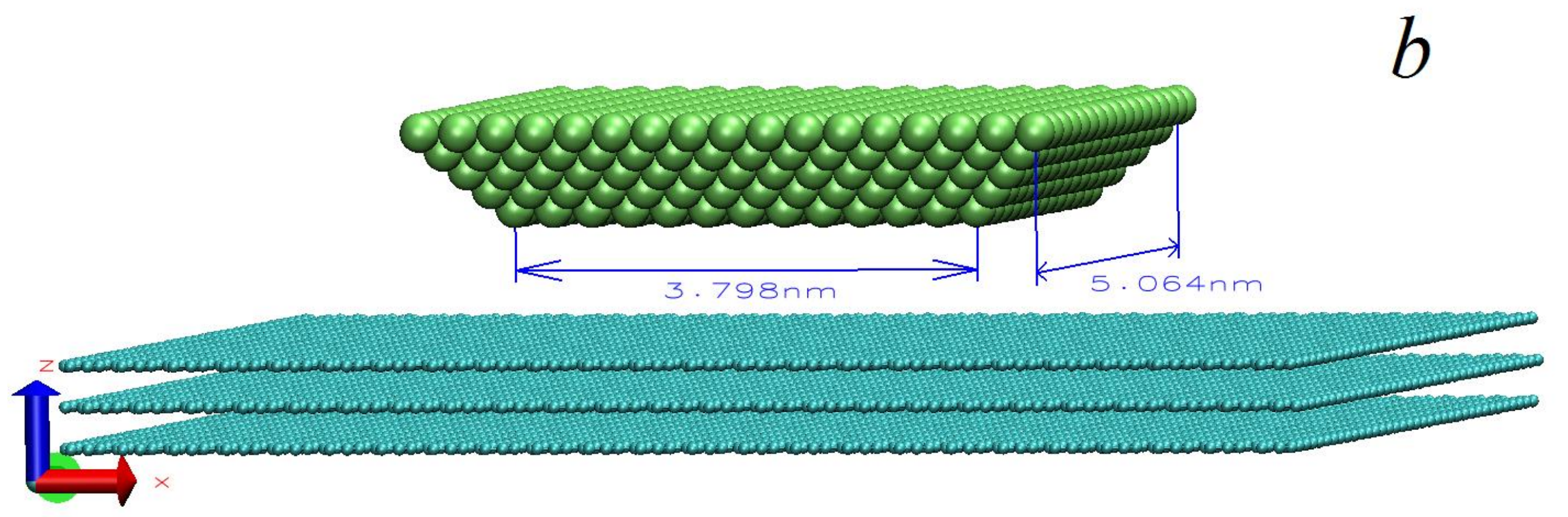}}
\caption{Top (\textit{a}) and perspective (\textit{b}) view of the initial atomic configuration of the studied system. Green and cyan balls correspond to tungsten and carbon atoms respectively (all snapshots in this work are produced with Visual Molecular Dynamics software~\cite{vmd}).}
\label{fig1}
\end{figure}

The graphitic sample consists of three graphene layers with AB stacking~(fig.~\ref{fig1}) which reflects $\alpha$ form of graphite. Armchair and zigzag graphene edges lie along $x$ and $y$ coordinate axes respectively and periodical boundary conditions are applied in the $xy$-plane. Each layer is composed of $24\times24$ honeycombs thus containing 3456 carbon atoms and the lengths along $x$ and $y$ directions are 10.082~nm and 8.731~nm respectively. To hold the sample in space, the bottom graphitic layer is rigid throughout the simulation.

Absolutely rigid square pyramidal nanoasperity (to which we also refer as to the tip) consists of five layers of atoms parallel to the $xy$-plane. Particles are arranged in a perfect bcc lattice with constant of 0.3165~nm and this corresponds to the crystal structure of tungsten\footnote{http://www.webelements.com}.The tapered form is provided by adding one atomic row in $x$ and $y$ directions per layer when moving from the bottom (which is the nearest to the sample part of the asperity) to the top of the tip. The bottom atomic layer exposes (001) crystallographic plane and has $13\times13$ atoms. The nanoasperity contains 1135 atoms and the total number of particles involved in the simulations is 11503. The dimensions of the tip are chosen to satisfy the fact that accordingly to the experiments the flake is assumed to attach to asperities on the tip with sizes of several nanometers (see HRTEM micrograph of the tungsten tip in fig.5.11 in~\cite{DienThes}). Strictly speaking, for completely realistic reproduction of the experimental conditions the bottom part of the nanoasperity in the model should be amorphous with appropriate parameters to reflect the presence of the oxide layer and the tip also should be able to deform. Nevertheless, as the first step towards solving the task we decided to study the system under mentioned approximations.

Covalent bonds between carbon atoms within two upper dynamic graphene layers are described by the Brenner potential. It has the following form~\cite{Brenn1990,REBO2002}
\begin{equation}
\label{eq1-brenner}
V_{B}=\sum_{i}\sum_{j>i}
[V^{R}(r_{ij}) - \overline{b}_{ij}V^{A}(r_{ij})].
\end{equation}
The functions $V^{R}(r_{ij})$ and $V^{A}(r_{ij})$ are pair-additive interactions that represent all interatomic repulsions (core-core, etc) and attraction from valence electrons, respectively. The quantity $r_{ij}$ is the distance between pairs of nearest-neighbor atoms $i$ and $j$, and $\overline{b}_{ij}$ is a bond order which reflects the kind of bond between atoms $i$ and $j$ and involves many-body effects necessary for proper description of bonding in hydrocarbons. In the current study expressions of a second-generation reactive empirical bond order (REBO) form of the potential~\cite{REBO2002} are used for pair-additive interactions. They give an improved fit to the elastic properties of diamond and graphite and more realistically model short-range hard wall repulsions as compared to the older version of the potential~\cite{Harri1991,Brenn1990}. For the sake of simplicity bond order function $\overline{b}_{ij}$ is chosen as in the first version of Brenner potential with parameters for potential II in ref.~\cite{Brenn1990}. The chosen simplified potential form implies that we do not intend to accurately simulate the in-plane behavior of graphene layers. The code from TREMOLO software~\cite{Grieb2007} is partly used in calculations of cubic splines and their derivatives in the bond order term, and the interactions from Brenner potential are computed using parallel algorithm presented in ref.~\cite{Cagla1999}.

For realistic modeling of processes related to cleavage of graphite the crucial role may play the proper description of the interactions between graphene layers~\cite{Ito2008,Kolmo2005}. A pairwise Lennard-Jones (LJ) potential can describe the overall cohesion between graphene layers, but it is much too smooth to describe variations in the relative alignment of adjacent layers. In this work we use registry-dependent interlayer potential (RDP) that can describe the corrugation in graphitic systems with reasonable accuracy~\cite{Kolmo2005}. It has the following form
\begin{eqnarray}
\label{eq2-kolmogorov}
V(\mathbf{r}_{ij},\mathbf{n}_{i},\mathbf{n}_{j})=
e^{-\lambda(r_{ij}-z_{0})}[C+f(\rho_{ij})+f(\rho_{ji})]
\nonumber\\
-A\left(\frac{r_{ij}}{z_{0}}\right)^{-6}.
\end{eqnarray}
The potential contains an $r^{-6}$ van der Waals (vdW) attraction and an exponentially decaying repulsion due to the interlayer wave-function overlap.
To reflect the directionality of the overlap the function $f$ is introduced which rapidly decays with the transverse distance $\rho$. The latter is defined via the distance $\mathbf{r}_{ij}$ between pairs of atoms $i$ and $j$ belonging to distinct layers and the vector $\mathbf{n}_{k}$ ($k=i,j$) which is normal to the $sp^{2}$ plane in the vicinity of atom $k$. In the present study $\mathbf{n}_{k}$ is computed as ``local'' normal, i. e. as average of the three normalized cross products of the displacement vectors to the nearest neighbors of atom $k$, and this corresponds to RDP1 in ref.~\cite{Kolmo2005}, where numerical values of the parameters can also be found. Here we mention only that for long-range vdW term the cutoff distance equal to $r_{\mathrm{c}}=2.7z_{0}=0.9018$~nm is used. The presence of normals in the RDP makes it in essence a many-body potential which requires much more computational effort as compared to simple pairwise potentials. In the current study interactions only between the adjacent layers are considered and they are computed using our specially developed parallel algorithm based on linked cell lists~\cite{Grieb2007,Rapap2004}.

The tip is assumed to interact only with the upper graphitic layer and interactions between the tungsten and carbon atoms are described via LJ pairwise potential
\begin{equation}
\label{eq3-tungsten-carbon}
V_{LJ}=\left\{\begin{array}{lr}
         4\varepsilon\left[
         \left(\frac{\sigma}{r}\right)^{12}-
         \left(\frac{\sigma}{r}\right)^{6}\right], & r < r_{\mathrm{c}}\\
         0, & r \geq r_{\mathrm{c}}
       \end{array}\right.,
\end{equation}
where $r$ is the distance between a pair of tungsten and carbon atoms, $\sigma=0.5z_{0}$, and the cutoff distance $r_{\mathrm{c}}$ is the same as for the RDP1. As the true value of forces acting between the tip and the surface are not firmly established, we investigate the behavior of the system for several values of $\varepsilon$, viz 0.1, 0.25, 0.5, 1, and 6~eV. The equations of motion are integrated using the leapfrog method~\cite{Rapap2004} with a time step $\Delta t=0.1$~fs. The heat is dissipated via the Berendsen thermostat coupled with two dynamic graphitic layers and implemented as in ref.~\cite{Grieb2007} with $\gamma=0.4$.

At this point it should be noted that several scenarios of the system's behavior can occur while moving the adhesive tip relatively to the sample. In the current study we consider two possible situations which are the contact of the nanoasperity with the substrate and the indentation of the graphitic sample. The former is observed when the tip is pulled toward the surface only prior to contact, after which it is retracted from the sample. The latter takes place when the nanoasperity is allowed to advance past the contact point and the sample is compressed by the tip. Extensive experimental and theoretical investigations of nanoindentation of a wide variety of materials have been performed during last two decades~\cite{Landm1989,Landm1990,Harri1992,Dedko2000,Schal2004,HeoSJ2005,Garg1998,Garg1999} and the following results pertaining to our study are worth mentioning. When highly adhesive tip approaches the substrate, jump-to-contact (JC) phenomena takes place~\cite{Landm1990,HeoSJ2005} which is manifested in upward ``jumps'' of surface atoms to wet the tip. In this case the contact between the tip and the sample is formed before tip's surface reaches the equilibrium position of the substrate surface. If the tip is further moved towards the sample the indentation occurs. In our computer experiments similar situations take place and we observe indentation of the sample although the bottom atoms of the tip do not reach the equilibrium height of the upper graphene layer. Additionally, when speaking about simulations where only the contact is considered, we will also sometimes refer to the movement of the tip as to indentation for brevity.

In the present study the indentation process proceeds as follows. After equilibration of the system during 10000 time steps at 298~K with the tip outside the range of interaction hung at 1.16~nm above the surface, the asperity was lowered towards the surface. Motion of the tip occurs by changing $z$-coordinates of tungsten atoms in increments of 0.01304~nm in simulations concerning indentation and of 0.03144~nm and 0.04716~nm when contact is considered. The entire system is equilibrated for 500 and 100 time steps in between displacements of the nanoasperity for indentation and contact respectively. Mentioned quantities correspond to indentation rates of 260.8~m/s, 3144~m/s, and 4716~m/s respectively. When contact is studied, the tip is not immediately withdrawn from the surface but it is hung for 2000 time steps in order to allow the formation of the contact between the proximal atomic layers of the two interfacing materials. In contrast, for indentation the tip is immediately pulled away from the surface after reaching the minimum height. The duration of simulations with slow and fast indentation rates is 10~ps and 2.5~ps respectively.

In experiments force-versus-distance curves are obtained, which reflect the changes of the normal force acting on the tip with a distance to the surface. In the present work this force is computed as the sum of $z$ components of forces acting on tungsten atoms from the graphitic sample, and they are averaged over the last 100 steps of the equilibration procedure in between displacements of the tip. Obtained in the simulations force-versus-distance curves as well as time dependencies of the normal force acting on the tip, potential energy of the system (per atom) and the interlayer energy of the upper two layers of the sample are presented in the next section.

\section{Results and discussion}
\label{results}

During 1~ps after the equilibration period of the simulations of indentation, when the forces between the tip and the sample are still zero, the average values of interlayer distance and the energy between the upper two dynamic graphene layers are about $0.336\pm0.004$~nm and $41.6\pm0.8$~meV respectively. These values differ from 0.334~nm and 48~meV computed for rigid layers using RDP~\cite{Kolmo2005} by about 1~\% and 15~\%. The discrepancy may be attributed to the finite cutoff distance used in the present study, thermal fluctuations of normals and the use of local normals instead of semilocal ones. Nevertheless, obtained quantities are very close to the experimental values~\cite{Kolmo2005}.

\begin{figure}[!]
\centerline{\includegraphics[width=0.51\textwidth]{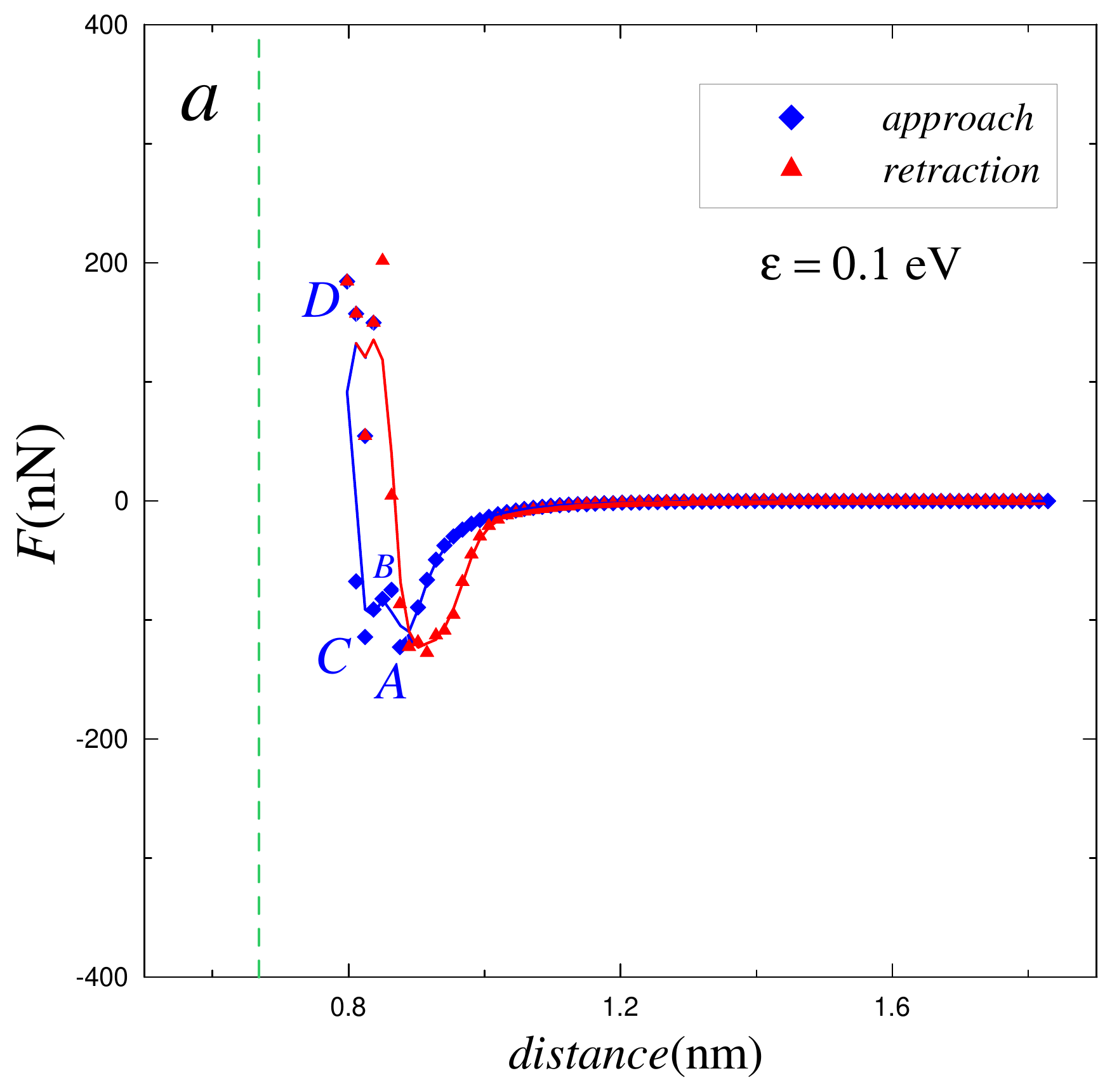}
\includegraphics[width=0.51\textwidth]{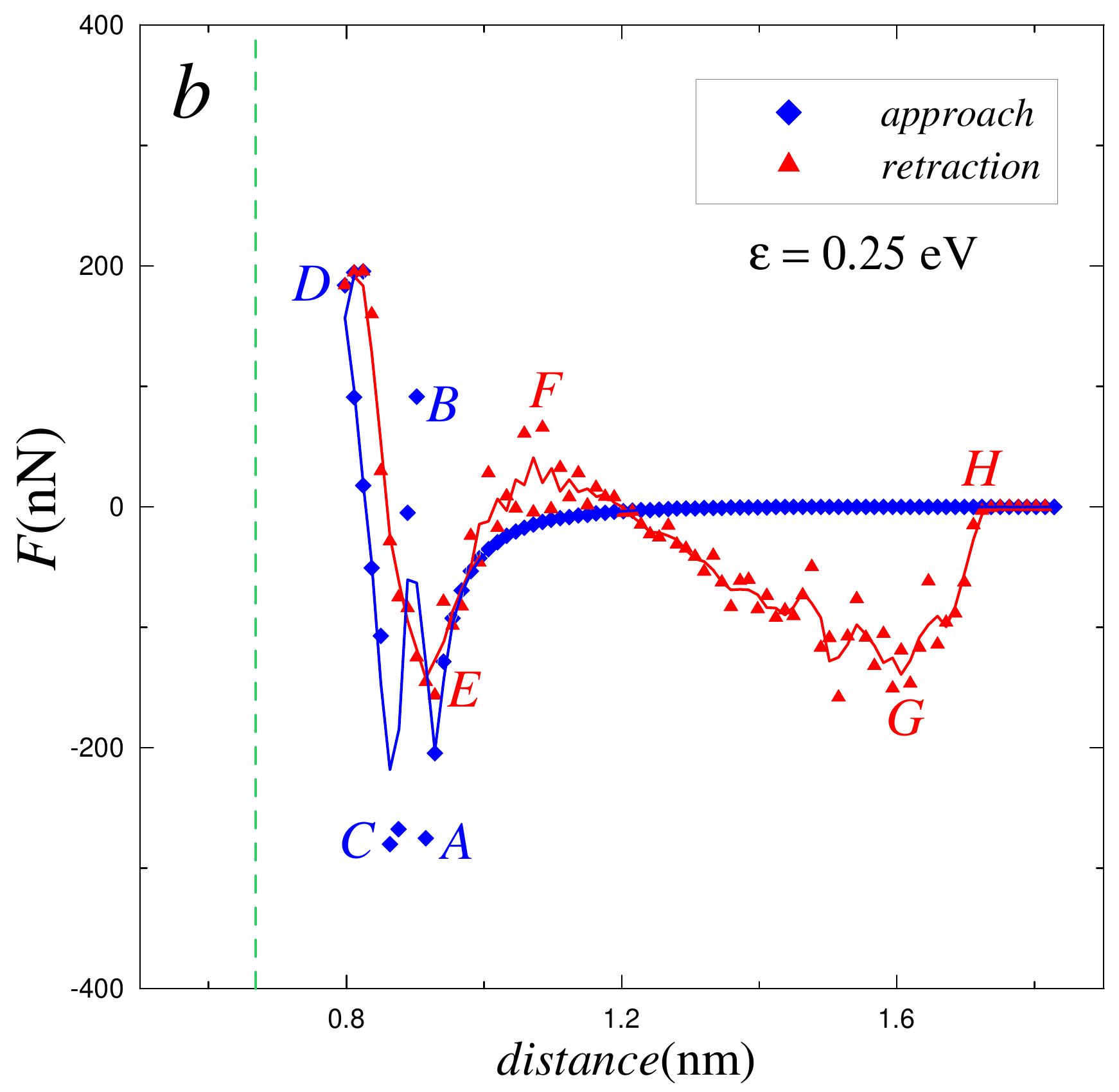}}
\centerline{\includegraphics[width=0.51\textwidth]{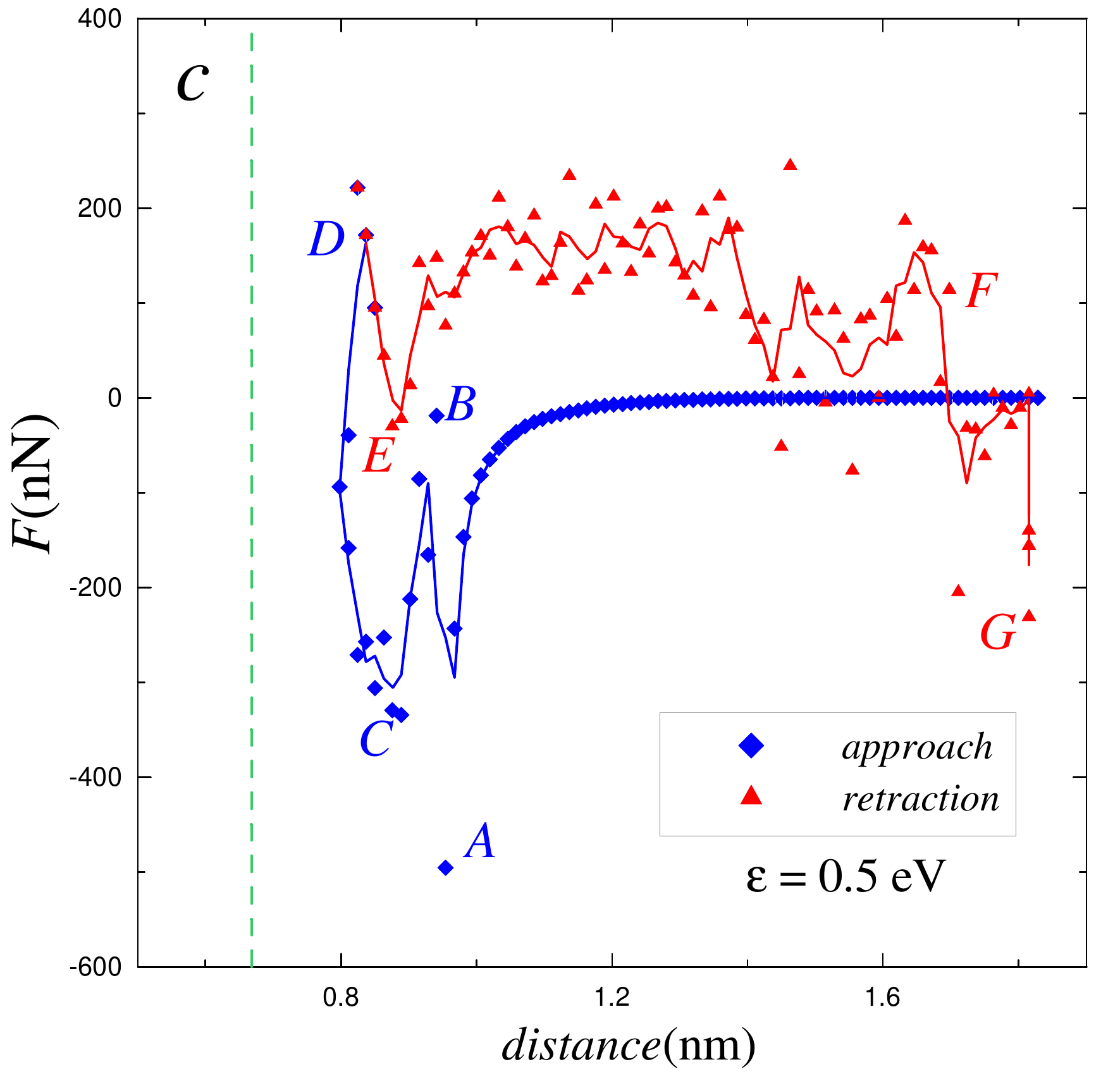}
\includegraphics[width=0.52\textwidth]{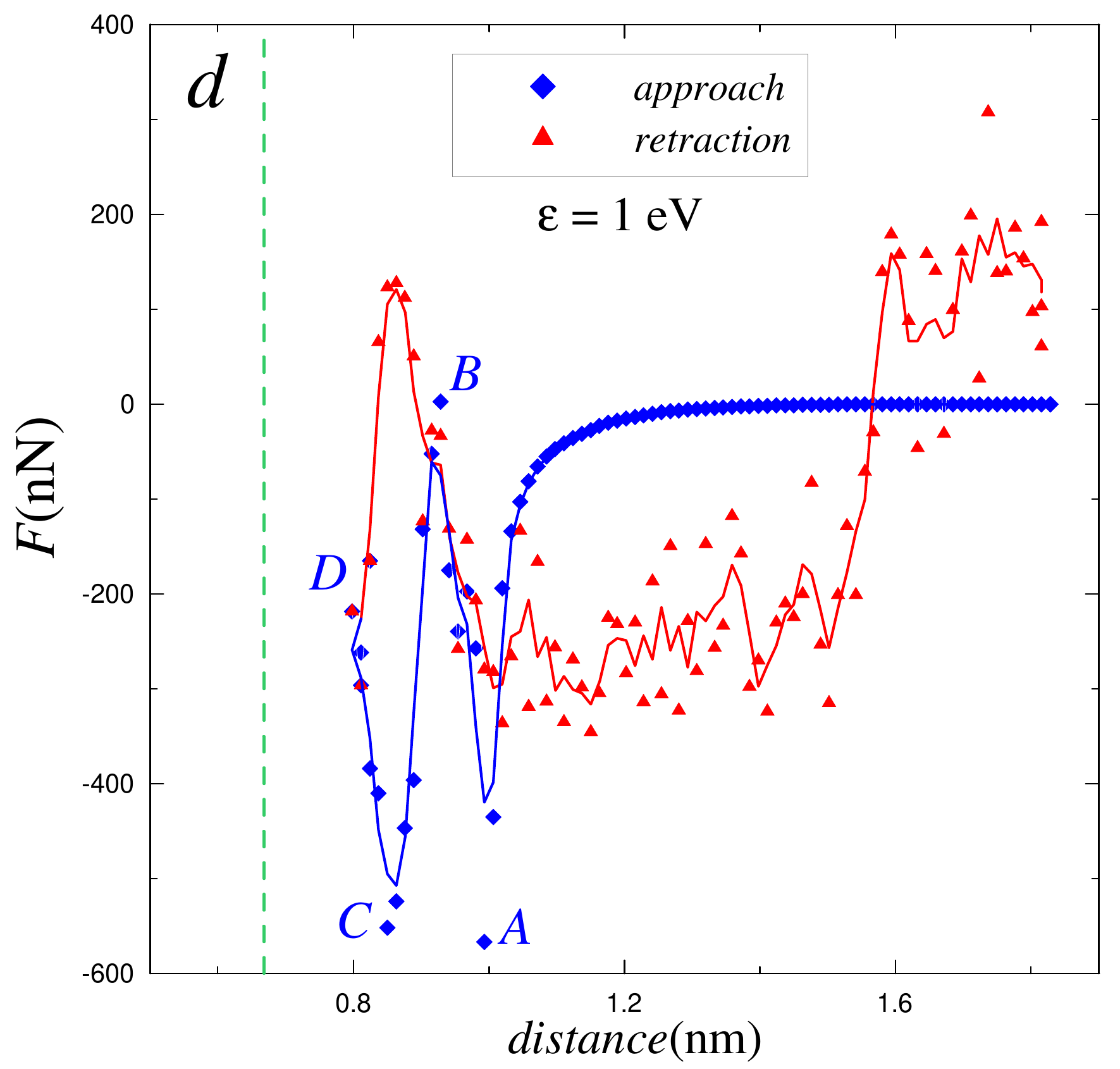}}
\caption{Normal force acting on the nanoasperity as it indents and then withdraws from the graphitic surface at rate of 260.8~m/s. Abscissa values correspond to the vertical distance between the rigid graphene layer and the bottom tungsten atomic layer. Solid lines depict the weighted average of measured data and are shown to follow the eye. Dashed line presents the equilibrium position of the upper carbon layer which is assumed to be 0.668~nm from the bottom layer.}
\label{fig2}
\end{figure}

\begin{figure}[!]
\centerline{\includegraphics[width=0.51\textwidth]{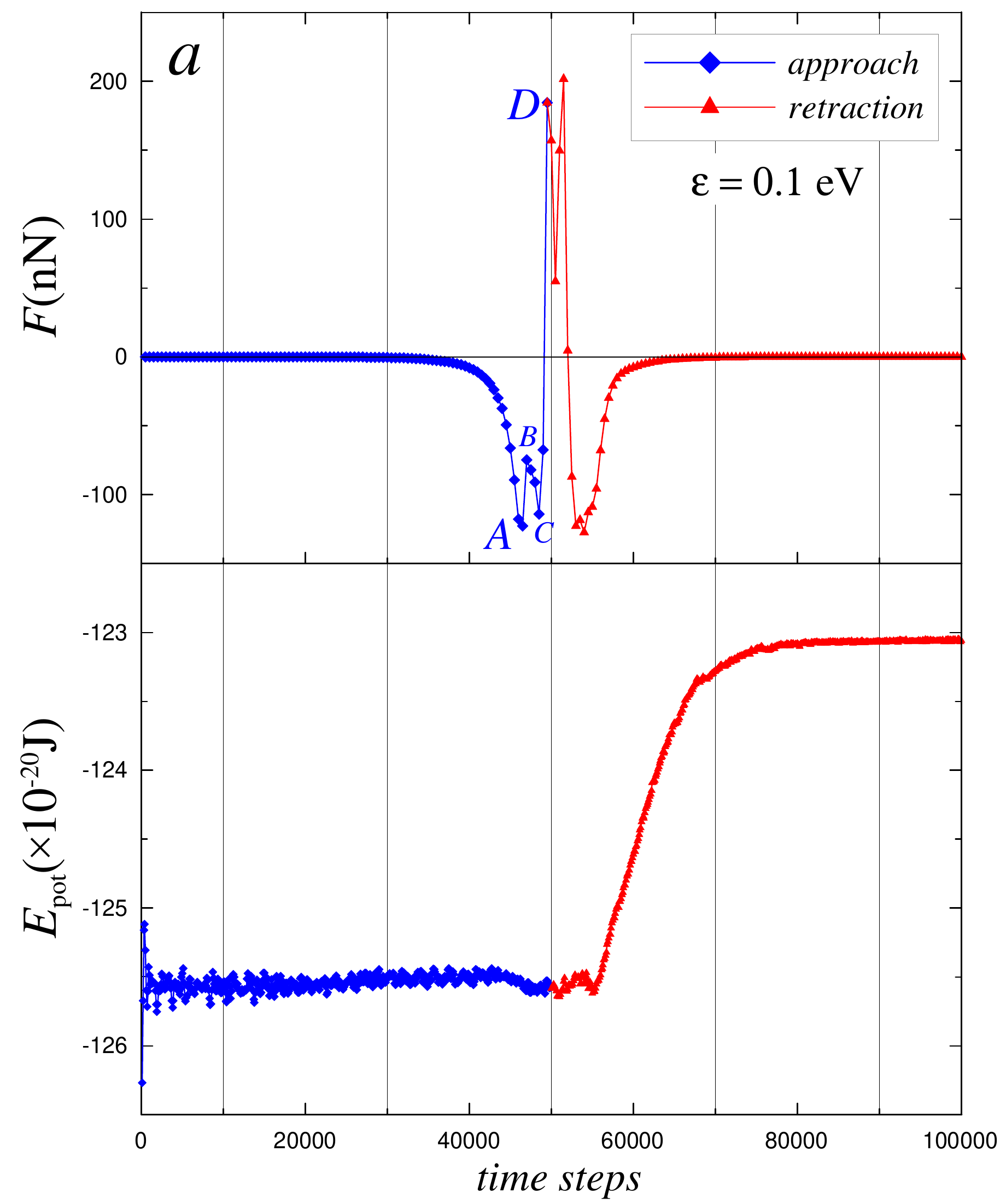}
\includegraphics[width=0.51\textwidth]{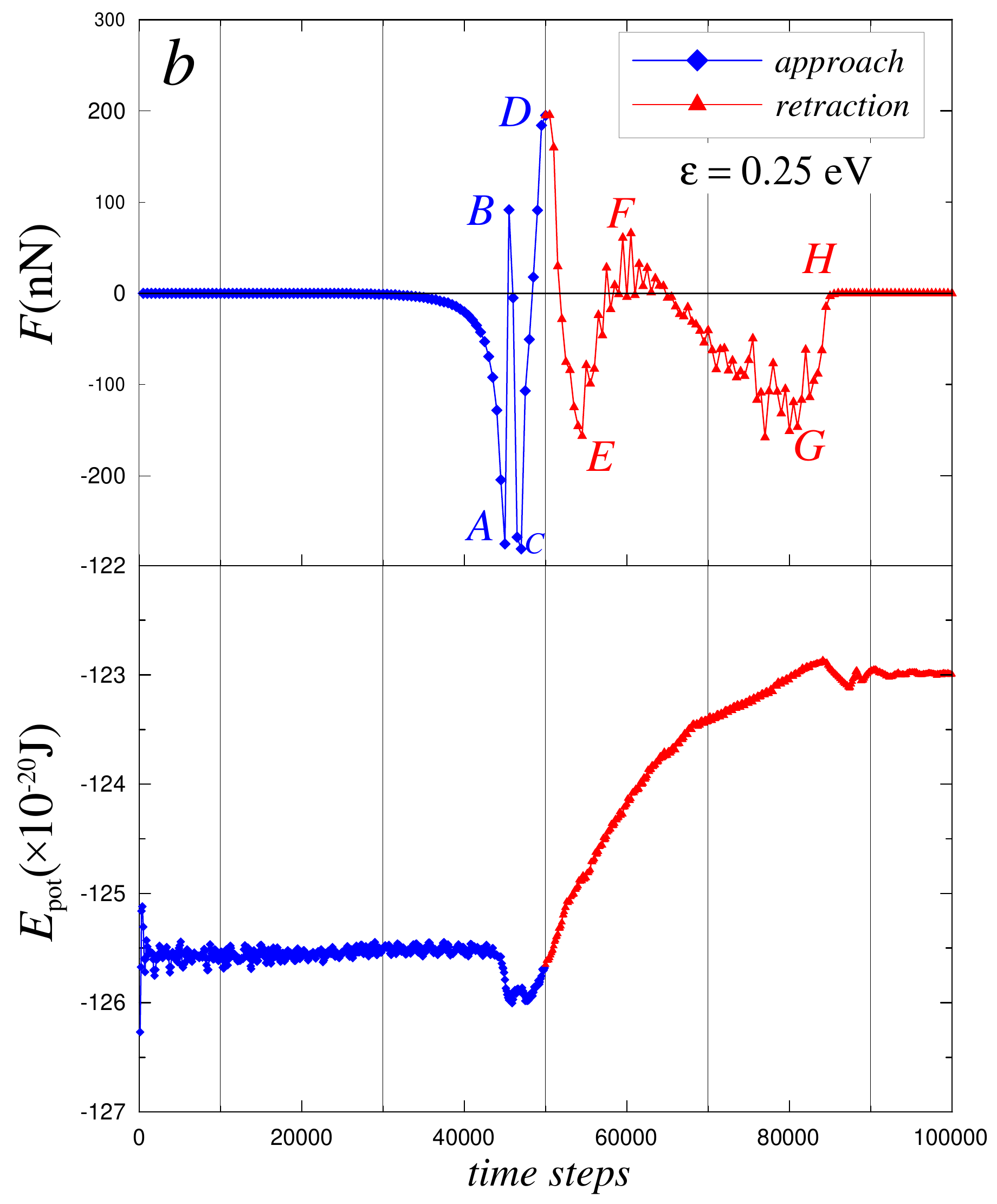}}
\centerline{\includegraphics[width=0.51\textwidth]{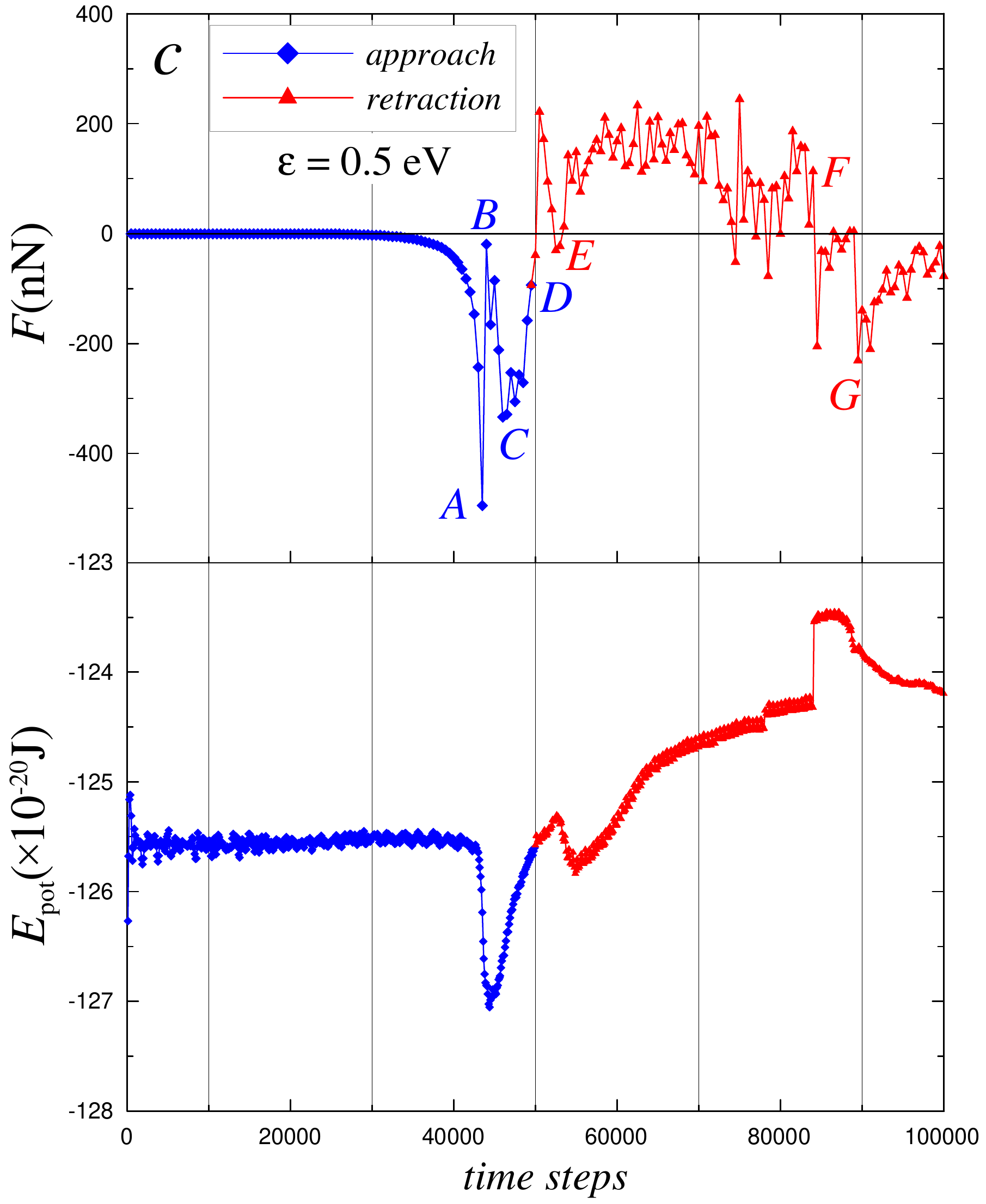}
\includegraphics[width=0.51\textwidth]{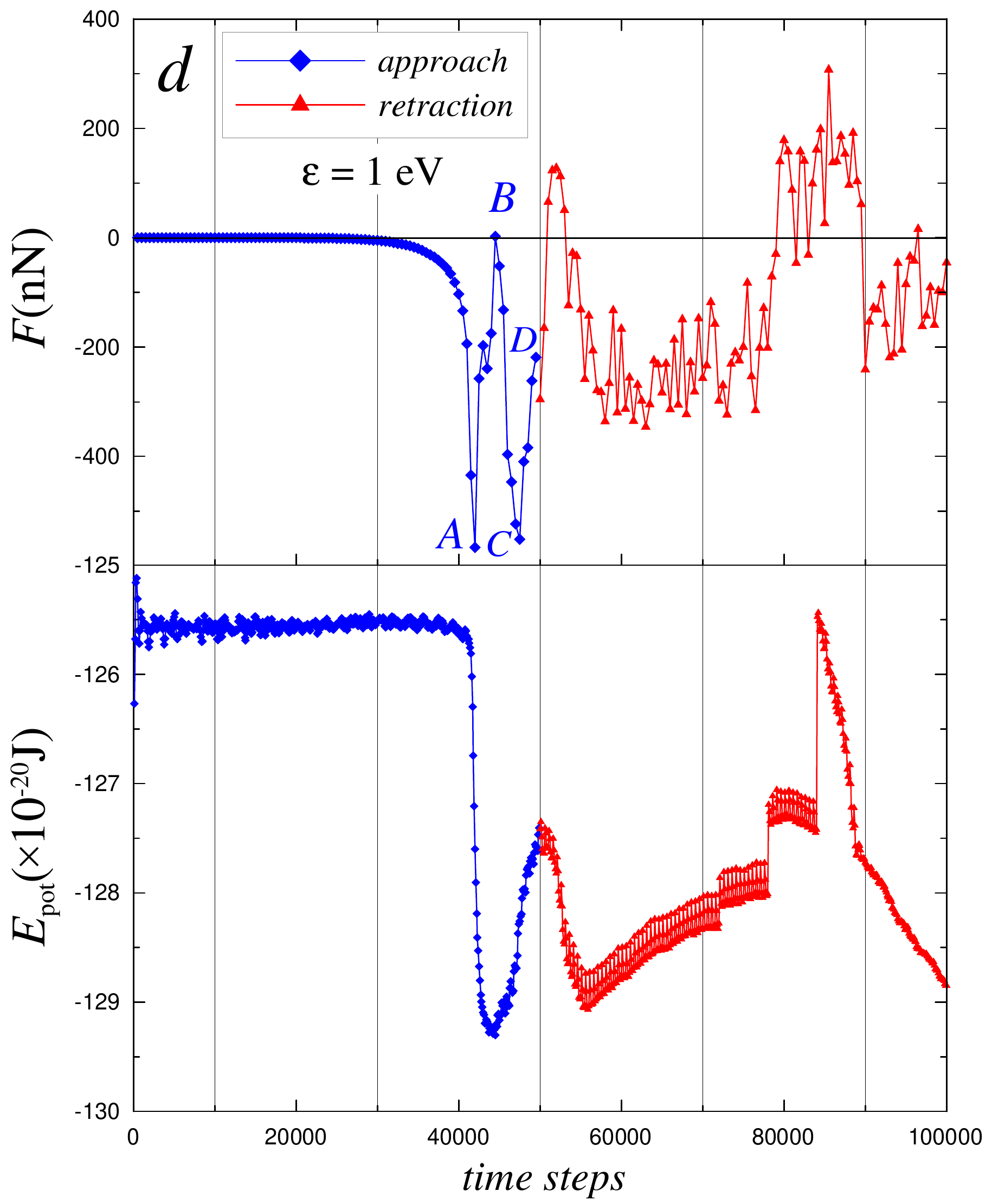}}
\caption{Time dependencies of the normal force acting on the tip and the potential energy of the system (per atom) for different magnitudes of interactions between the tip and the substrate.}
\label{fig3}
\end{figure}

\subsection{Indentation}

\begin{figure}[htb]
\centerline{\includegraphics[width=0.52\textwidth]{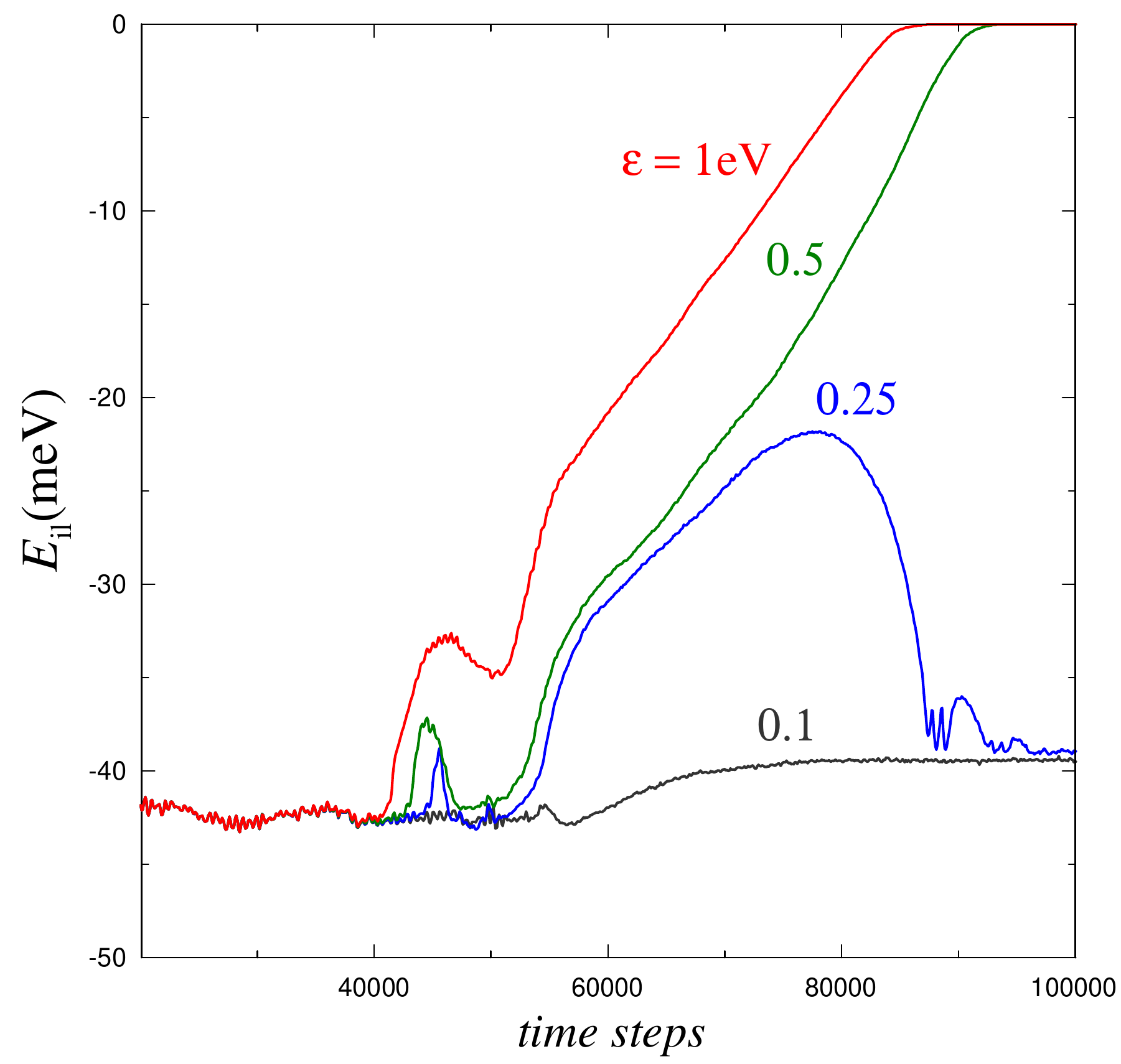}}
\caption{Time dependencies of the energy between the upper two graphene layers (after equilibration period) for different magnitudes of tip-sample interactions.}
\label{fig4}
\end{figure}

Fig.~\ref{fig2} and fig.~\ref{fig3} show force-versus-distance curves and time dependencies of the normal force $F$ acting on the nanoasperity and potential energy $E_{\mathrm{pot}}$ of the system (per atom) obtained in simulations corresponding to the indentation for different values of $\varepsilon$ in eq.(\ref{eq3-tungsten-carbon}). Following an initial slow variation of the force between the graphite substrate and the tungsten nanoasperity as the latter is being pulled toward the surface, the onset of an instability is observed, signified by a sharp increase in the attraction between the two. This is accompanied by a decrease in the potential energy of the system (see fig.~\ref{fig3}) which is larger the stronger tip-sample interactions are. The maximum attraction (point \textit{A} in fig.~\ref{fig2} and fig.~\ref{fig3}) corresponds to a JC phenomenon which occurs via a fast process where carbon atoms under the asperity displace toward it in a short time span of $\sim0.5$~ps. JC is associated primarily with a tip-induced sample deformation~\cite{Landm1990}, which begins when the distance between the proximal atomic layers of the two interfacing materials is in the range of approximately 0.19~nm for the smallest $\varepsilon$ and 0.33~nm for the largest one. It is further evidenced by time dependencies of interlayer energy $E_{\mathrm{il}}$ between the upper two graphene layers shown in fig.~\ref{fig4}, where sharp peaks or fast increase of $E_{\mathrm{il}}$ are observed in between 40000 and 50000 time steps for $\varepsilon=0.25, 0.5$, and 1~eV respectively. JC leads to the collision of carbon atoms with absolutely rigid nanoasperity, which causes a sudden drop of attraction (point \textit{B} in fig.~\ref{fig2} and fig.~\ref{fig3}) the magnitude of which is greater for stronger tip-substrate interactions. Further advancement of the tip results in the decrease of repulsion, and after reaching the point \textit{C} a new dramatic increase in force is observed (\textit{CD} segment of the curves) indicating the repulsive wall region which corresponds to indentation of the sample~\cite{Schal2004,HeoSJ2005}.

In some works concerning molecular dynamics of nanoindentation~\cite{Landm1990,Harri1992,Garg1998,Garg1999} the conditions of constant temperature are maintained. However, in the present study of indentation in spite of strong coupling of atoms to the thermostat the heat has not enough time to completely dissipate during equilibration in between the tip displacements. The heating of the sample during indentation was in the range from 30~K to 70~K for the smallest and the largest values of $\varepsilon$ respectively.

It should be mentioned that for $\varepsilon=0.1$~eV the JC phenomenon is very weakly manifested in fig.~\ref{fig2}\textit{a} or fig.~\ref{fig3}\textit{a}, and in fig.~\ref{fig4} it is not apparent at all. Nevertheless, its presence is confirmed by animations obtained during the simulations and it is also evident from the repulsive wall region in fig.~\ref{fig2}\textit{a} and fig.~\ref{fig3}\textit{a} indicative of indentation and therefore of the direct interfacial contact. The latter would not be possible without JC, as the minimum distance between the bottom atoms of the nanoasperity and the equilibrium position of the upper carbon layer in all simulations of indentation is about 0.12~nm.

Lifting of the tip after indentation results in enhanced adhesion between the tip and the surface, which is evidenced by hysteresis in the force curves for all magnitudes of tip-substrate interactions.
Let us consider withdrawal of the nanoasperity for each value of $\varepsilon$ in more detail. Retraction begins after 50000 time steps in all simulations pertaining to indentation.

\begin{figure}[htb]
\centerline{\includegraphics[width=0.5\textwidth]{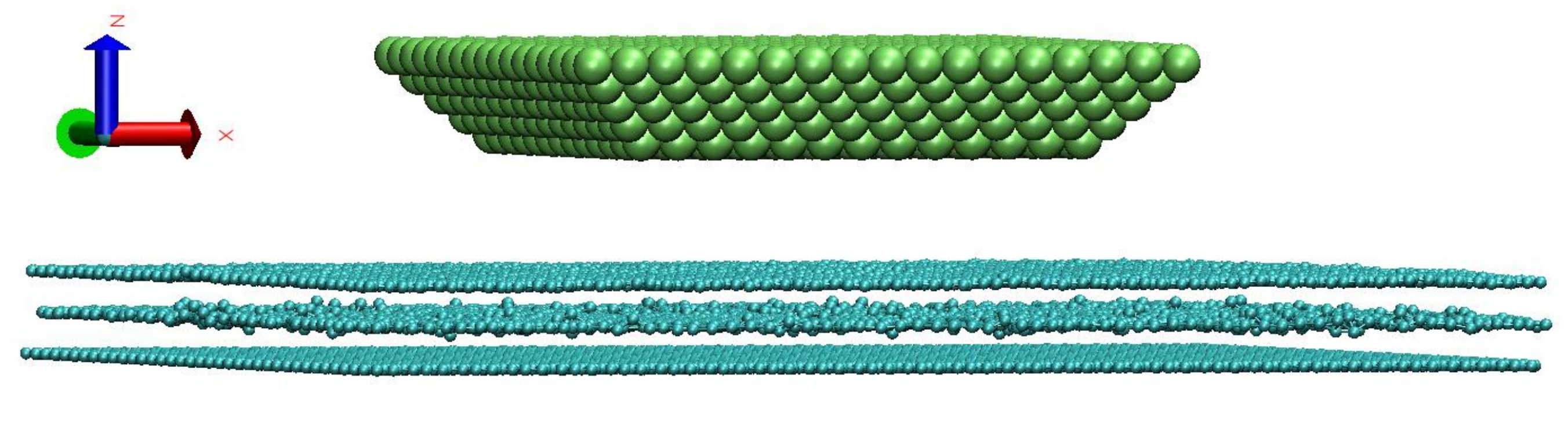}}
\caption{System at the end of the simulation of indentation with $\varepsilon=0.1$~eV. Some carbon atoms in the middle layer left $sp^{2}$ plane and caused the formation of the defective atomic structure.}
\label{fig5}
\end{figure}

For $\varepsilon=0.1$~eV force-displacement curve in fig.~\ref{fig2}\textit{a} during separation exhibits almost the same form as during loading, has a monotonic form with a slight hysteresis caused by adhesion of atoms, and there are no significant changes in interlayer energy in fig.~\ref{fig4}. These facts and the eventual approach of the force to zero indicate the absence of cleavage of the sample for the considered tip-sample interactions. As a video animation of the computer experiment shows, the increase in potential energy and in $E_{\mathrm{il}}$ during retraction is caused by mechanical instability in the middle graphene layer, which occurs as a consequence of a perturbation from the tip and results in the formation of the defective structure of this layer. The latter can be observed in fig.~\ref{fig5}.

\begin{figure}[htb]
\centerline{\includegraphics[width=0.51\textwidth]{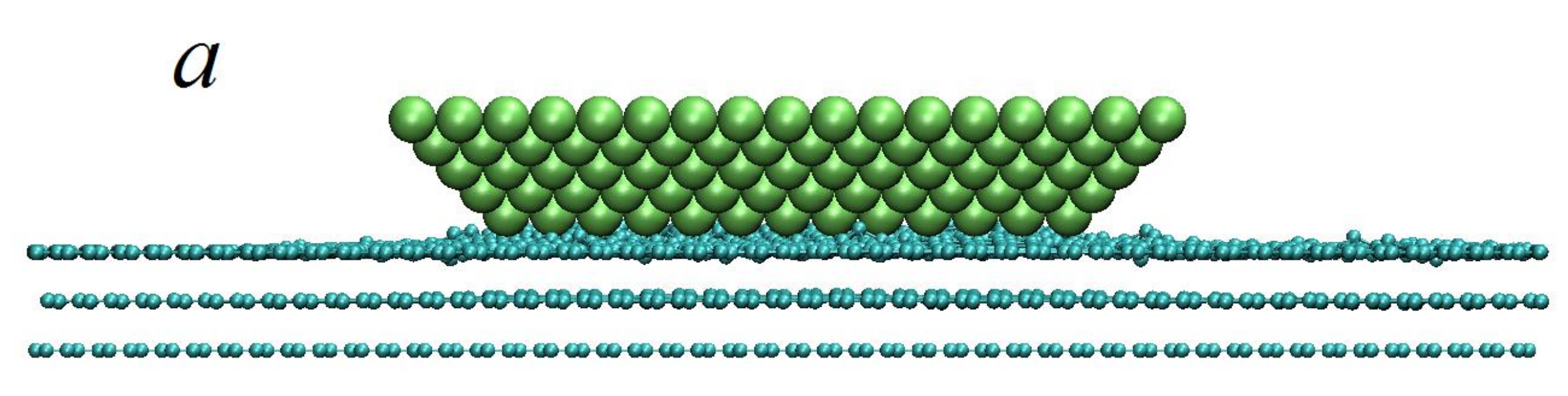}
\includegraphics[width=0.51\textwidth]{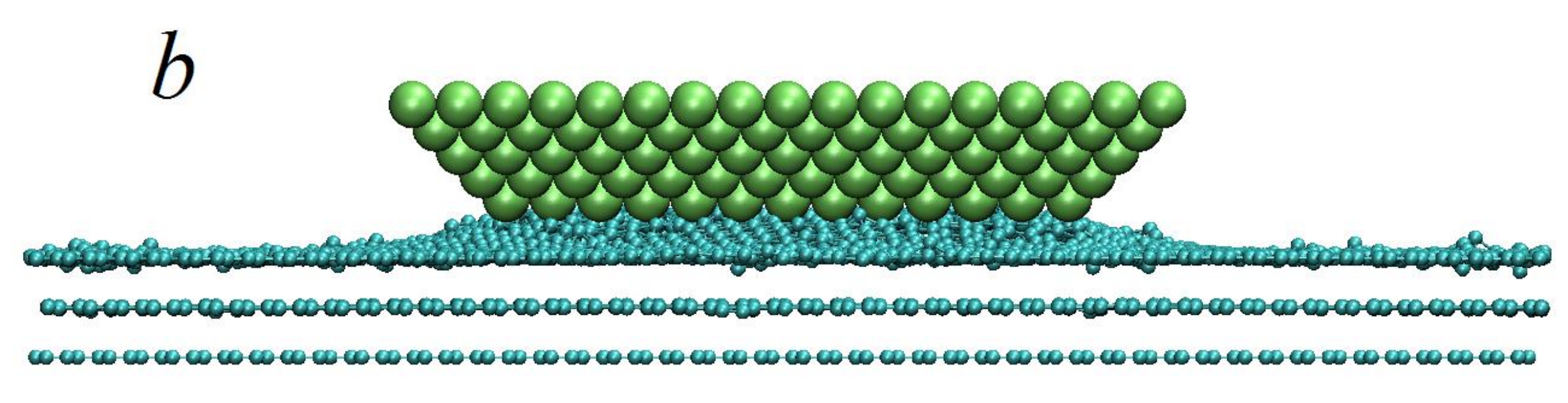}}
\centerline{\includegraphics[width=0.51\textwidth]{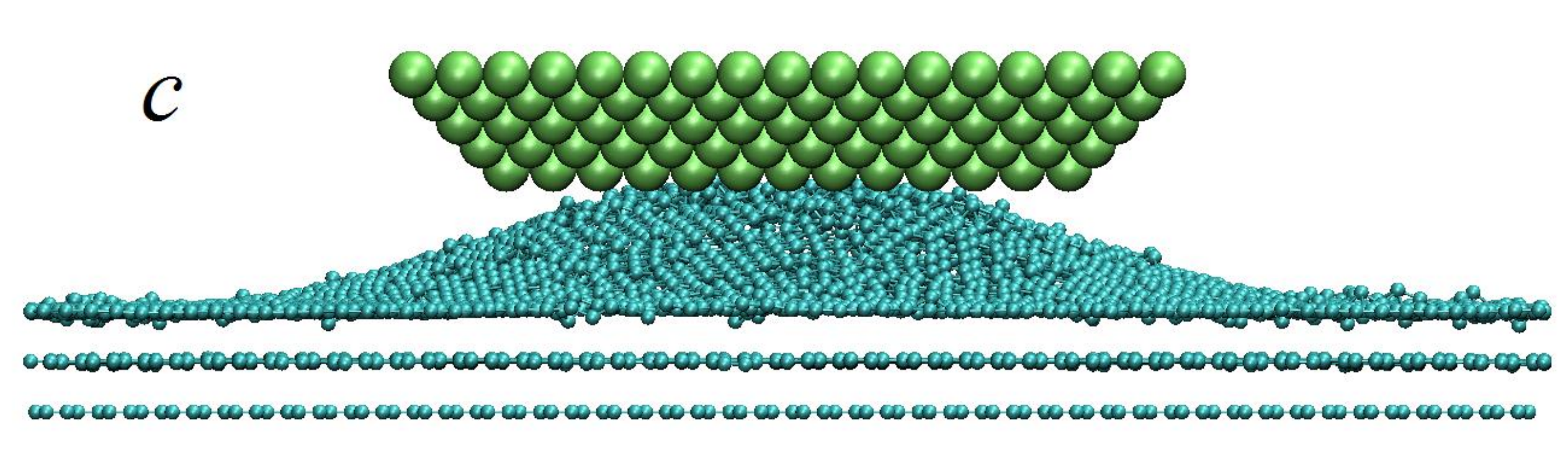}
\includegraphics[width=0.51\textwidth]{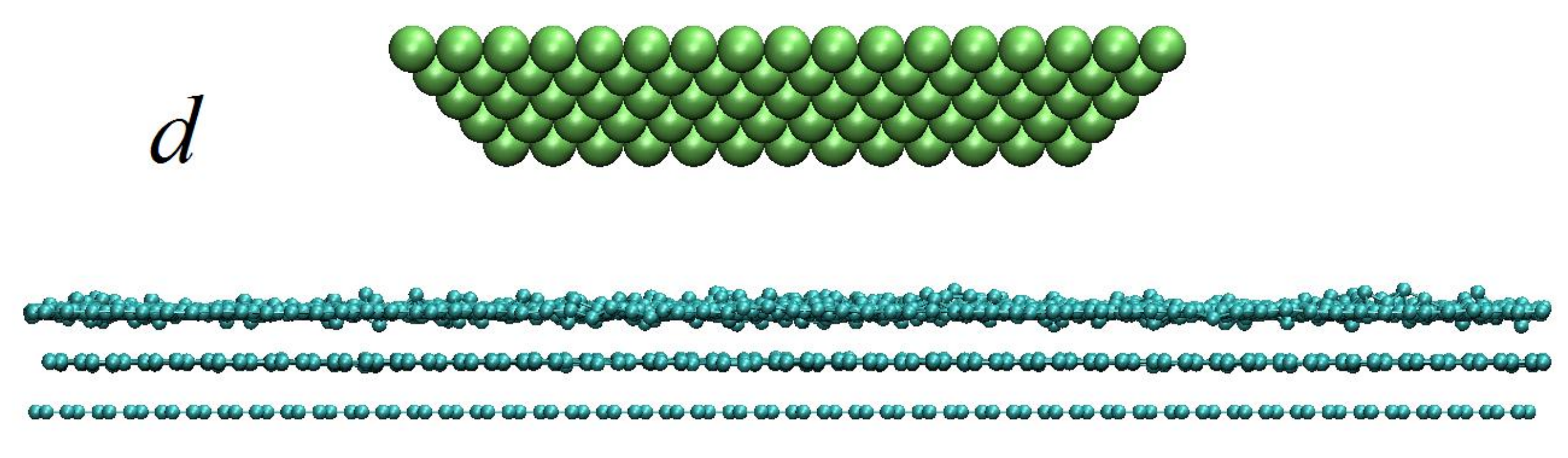}}
\caption{Instantaneous atomic configurations during indentation with $\varepsilon=0.25$~eV corresponding to the following regions in fig.~\ref{fig2}\textit{b} and fig.~\ref{fig3}\textit{b}: (a) point \textit{E}, (b) point \textit{F}, (c) middle of segment \textit{GH}, (d) final state. The tip ``loses'' carbon atoms and defective structure is formed in the upper carbon layer.}
\label{fig6}
\end{figure}

The withdrawal part of force dependencies in fig.~\ref{fig2}\textit{b} and fig.~\ref{fig3}\textit{b} for $\varepsilon=0.25$~eV can be subdivided into four main parts (fig.~\ref{fig6} shows several corresponding snapshots of the system). Segment \textit{DE} reveals the increase of attraction while the tip is pulled away from the substrate and reflects the tendency of carbon atoms to withstand the retraction. However, after point \textit{E} the attraction decreases and the repulsion becomes dominating. This suggests that carbon atoms in the upper layer tend to move upward (thus closely approaching the tip and increasing repulsion force) and the trend to exfoliation of the upper layer can be observed from the increase of $E_{\mathrm{il}}$ in fig.~\ref{fig4}. Nevertheless, the used value of  $\varepsilon$ is not enough to cleave the upper layer, so after reaching the distance of about 1.2~nm in segment \textit{FG} attraction becomes dominating once more, and in segment \textit{GH} the tip ``loses'' carbon atoms which return to the equilibrium vertical position of the upper layer. This is also corroborated by a fast drop in $E_{\mathrm{il}}$ after 80000 time steps and the eventual approach of the force to zero value. The described results are similar to the experimentally observed behavior of a FFM tip when it interacts with a surface contaminated with liquid molecules. The tip must be pulled a certain distance, the break-free length, to break free from the meniscus of liquid molecules~\cite{Mate2002}. In our case, however, the role of liquid molecules play carbon atoms and lamellar structure causes the repulsive segment \textit{EF} which is absent in the mentioned experiment. The ultimate increase in the potential energy and in $E_{\mathrm{il}}$ as compared to the initial values is a result of the formation of the defective structure in the upper layer as can be seen in fig.~\ref{fig6}\textit{d}. This is distinct from the previous case, where defects were generated in the second layer.

\begin{figure}[htb]
\centerline{\includegraphics[width=0.51\textwidth]{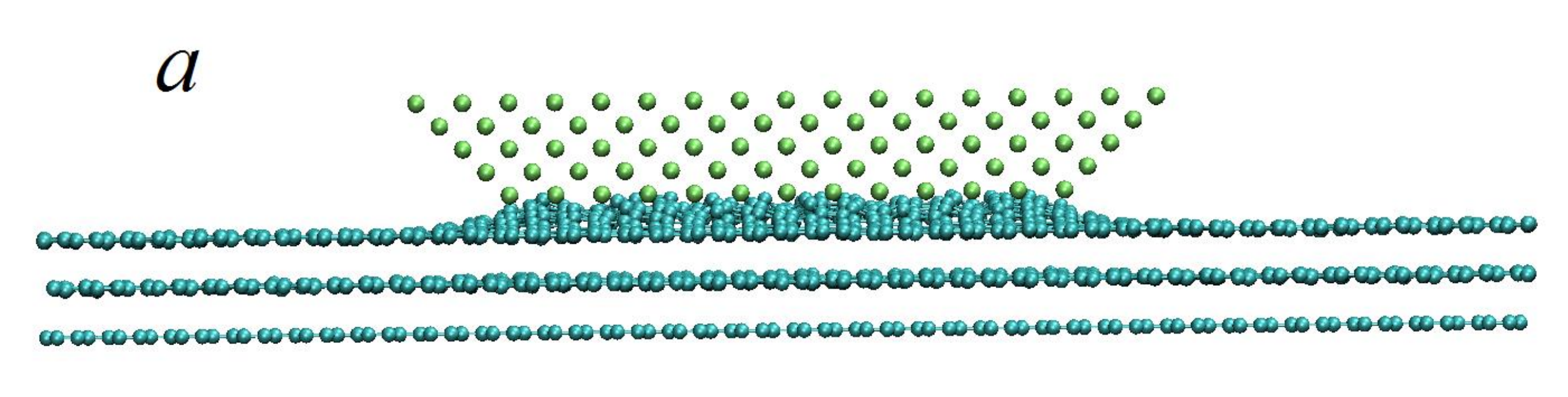}
\includegraphics[width=0.51\textwidth]{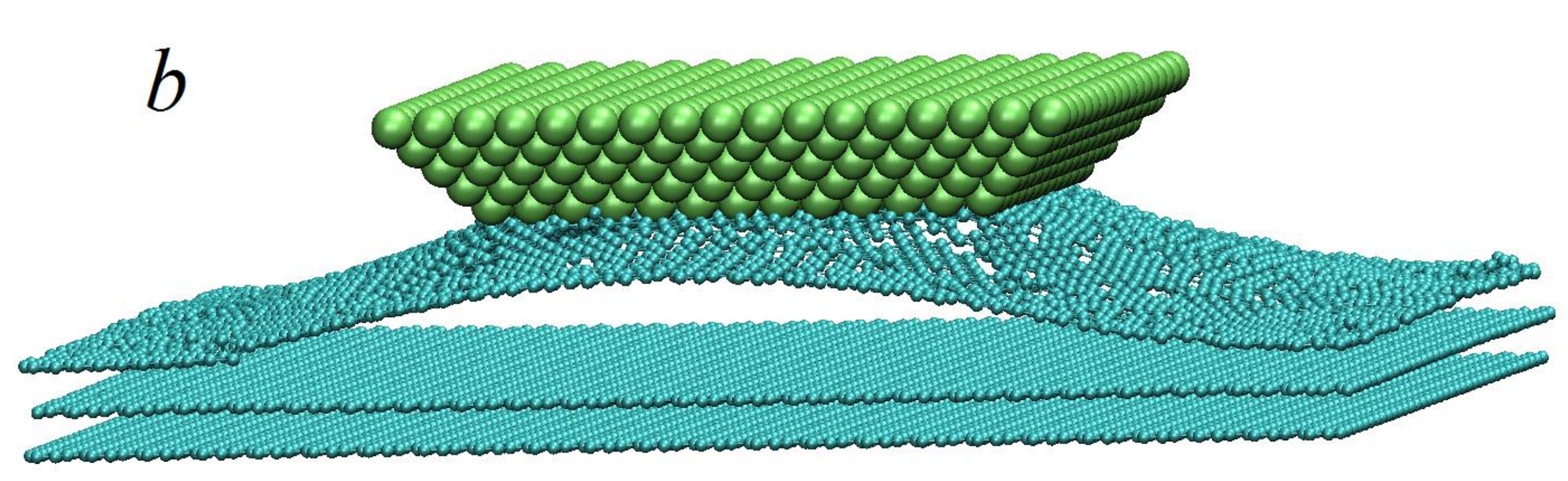}}
\centerline{\includegraphics[width=0.51\textwidth]{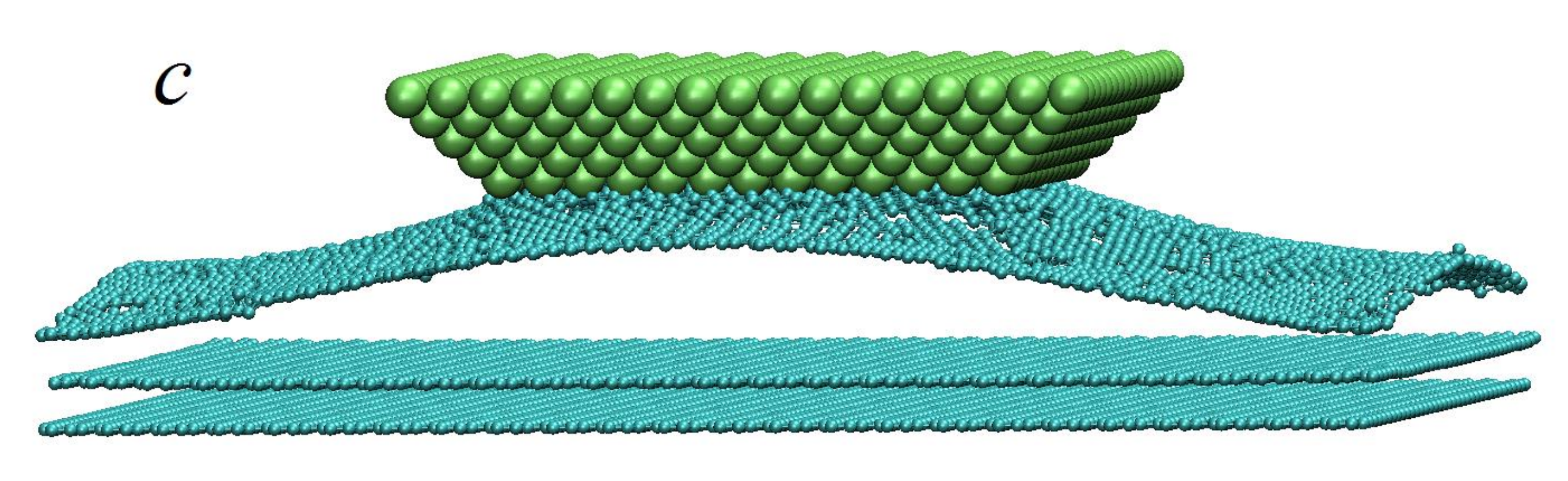}
\includegraphics[width=0.51\textwidth]{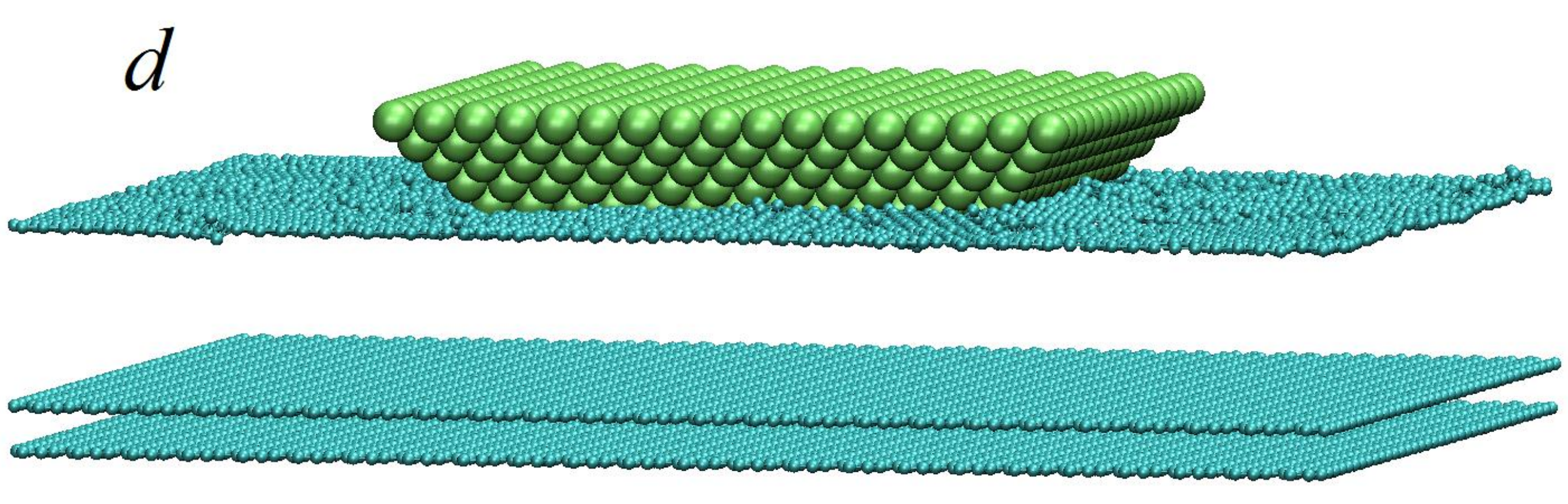}}
\caption{Snapshots of the system during indentation with $\varepsilon=0.5$~eV corresponding to the following parts of fig.~\ref{fig2}\textit{c} and fig.~\ref{fig3}\textit{c}: (a) JC point \textit{A} (radius of tungsten atoms is diminished for clarity), (b) point \textit{F}, (c) point \textit{G}, (d) exfoliation of the upper layer at the end of the simulation.}
\label{fig7}
\end{figure}

Value of $\varepsilon=0.5$~eV is enough for complete exfoliation of the upper layer as fig.~\ref{fig2}\textit{c}, fig.~\ref{fig3}\textit{c}, and fig.~\ref{fig7} suggest. After moderate increase of attraction till point \textit{E} the force is mainly repulsive before point \textit{F} is reached. This exhibits the tendency of carbon atoms to push the tip upwards and hence to move in this direction, thus stimulating the exfoliation of the upper layer. Sudden change of repulsion to attraction after point \textit{F} in fig.~\ref{fig2}\textit{c} and fig.~\ref{fig3}\textit{c} is indicative of the final stage of exfoliation, where forces between graphene sheets at their boundaries should be overcome and the buckling of the upper layer is observed (see fig.~\ref{fig7}\textit{c}). The ultimate configuration shows the completely removed upper layer (fig.~\ref{fig7}\textit{d}) corresponding to zero interlayer energy in fig.~\ref{fig4}.

The nonmonotonic behavior of the force-versus-distance relationship during separation after indentation is apparent for larger values of $\varepsilon$. It was also observed in simulations of nanoindentation of metals~\cite{Landm1990,Dedko2000} and was associated with atomic structural transformations during elongation of the connective neck which occurs in metallic systems. In our case scattering of data points can be attributed to a relatively large displacement step of the tip which leads to considerable impacts of carbon atoms with nanoasperity, as the stress induced in graphene sheet by tip movement has enough time to relax in between the displacements of the nanoasperity, in contrast to fast indentation as will be shown later. The absolute rigidity of the tip may also contribute to such a behavior as the tungsten atoms cannot response to collisions by deformation to soften the hits thus diminishing the scattering. In some works additional averaging is performed on the force-displacement curves to filter out the noise from thermal vibrations~\cite{Garg1998,Garg1999}. This has not been carried out in the present study.

\begin{figure}[htb]
\centerline{\includegraphics[width=0.5\textwidth]{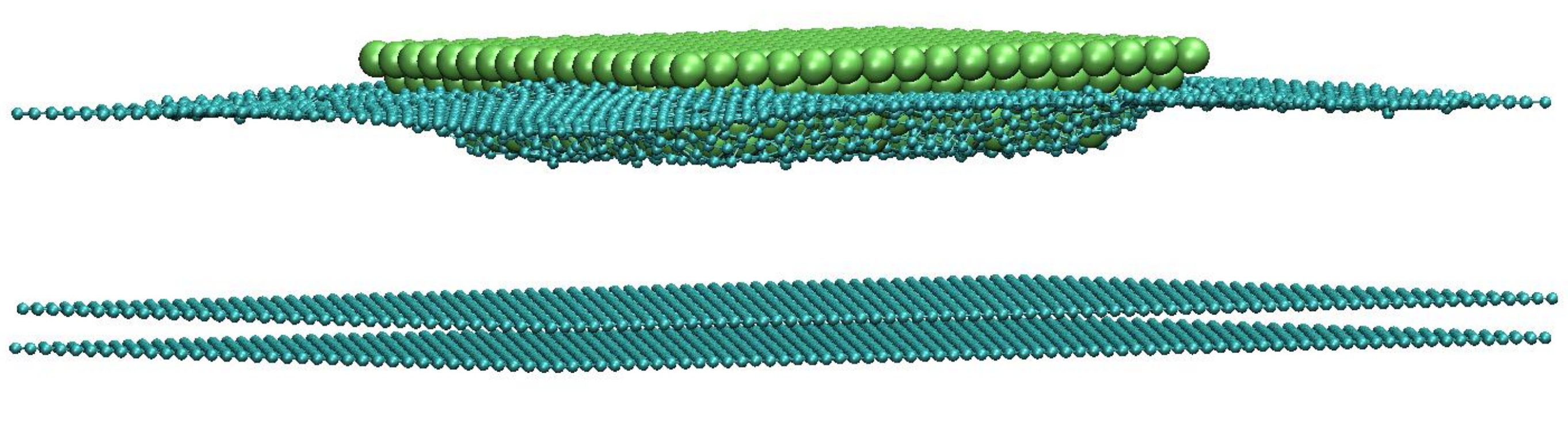}}
\caption{Final configuration of the system after indentation for $\varepsilon=1$~eV.}
\label{fig8}
\end{figure}

For $\varepsilon=1$~eV the exfoliation also takes place. Although curves in fig.~\ref{fig2}\textit{d} and fig.~\ref{fig3}\textit{d} have distinct from the previous case shapes, they can be interpreted in a similar way. Here we mention that such a large attraction causes considerable adhesion-induced wetting of the edges of the tip by carbon atoms such that atoms attach to the sides of the nanoasperity as can be seen in fig.~\ref{fig8}.

Thorough inspection of the cleaved layer for both considered cases reveals its highly defective structure. It has relatively large regions with rearrangement of atoms in configurations with more than three nearest neighbors in contrast to the initial graphitic honeycomb lattice. These structural transformations are responsible for changes in potential energy in fig.~\ref{fig3}\textit{c} and fig.~\ref{fig3}\textit{d} (changes in interlayer energy do not considerably influence the total potential energy of the system, because $E_{\mathrm{il}}$ is smaller by about 3 orders of magnitude than $E_{\mathrm{pot}}$). However, experiments show that defects in graphene can exist without considerable structural rearrangements of surrounding regions~\cite{Gass2008}. Transformations observed in the simulations cannot be the consequence of heating as it was rather low. Therefore, we conclude that the old version of the bond order function in Brenner potential used in the current work is not capable of proper description of defects in graphene. This also may be the reason for generation of defects for $\varepsilon=0.1$ and 0.25~eV mentioned above.

\begin{figure}[!]
\centerline{\includegraphics[width=0.5\textwidth]{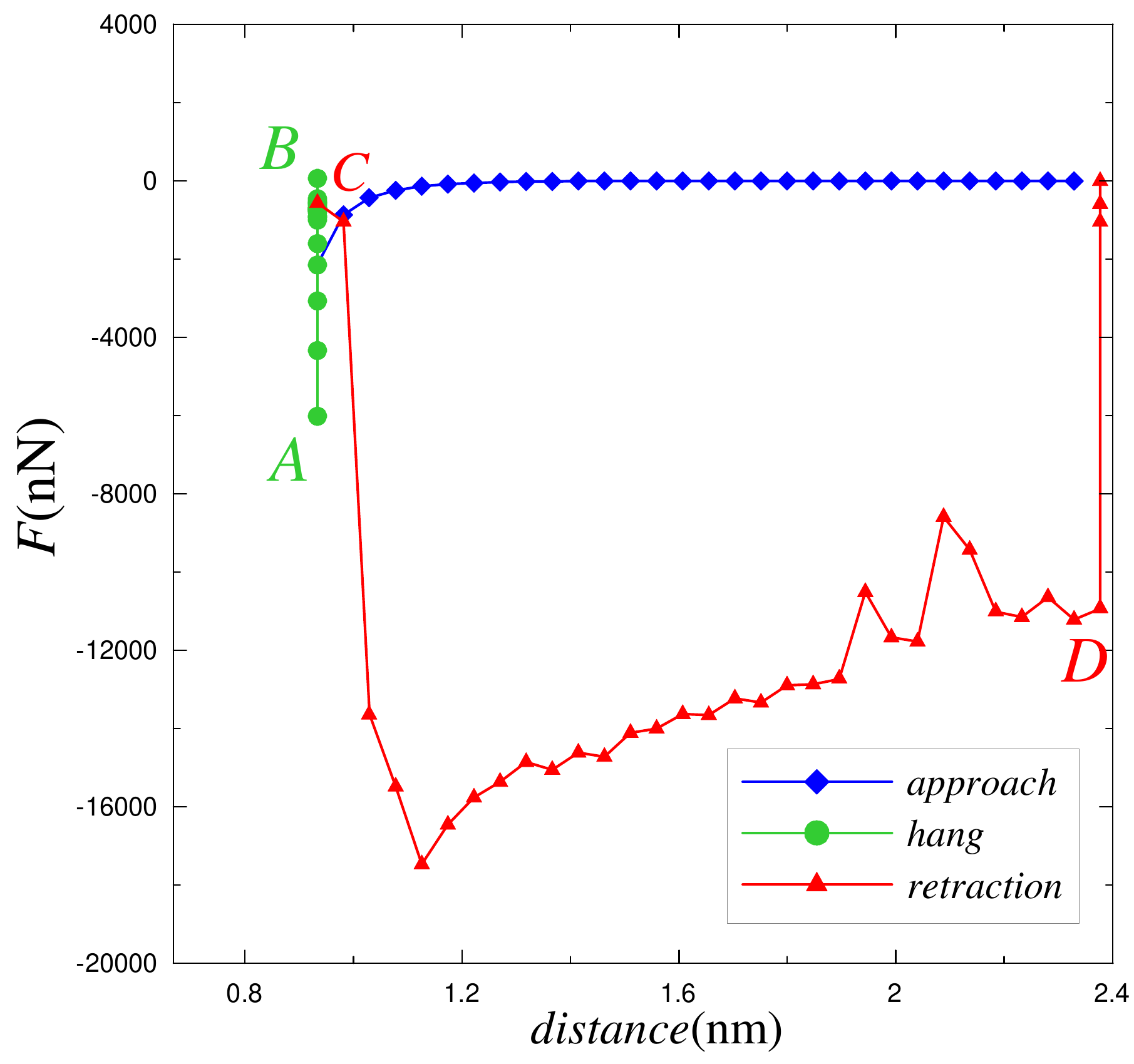}}
\caption{Force-displacement curve obtained during simulation of contact. The equilibrium position of the upper graphitic layer coincides with the minimum abscissa value of the plot.}
\label{fig9}
\end{figure}

\subsection{Contact}

In order to explore the possibility of the formation of a flake attached to the nanoasperity a second series of simulations has been carried out where contact has been considered. In these computer experiments the tip is moved at high rates (given in section~\ref{model}) and is not allowed to compress the substrate, but after the contact formation is immediately pulled away from the sample. Considerably higher rates comparatively to indentation are required to provide the conditions where the excess heat has not enough time to dissipate and to reach high value of temperature enough for contributing in breaking of covalent bonds in a graphene sheet.
However, as computer experiments have shown, none combination of values of the parameters mentioned in section~\ref{model} produced the desired result. So we carried out an additional simulation where we enlarged the initial distance between the upper carbon layer and the bottom tip layer by about 0.5~nm, and used the values of 4814~m/s and 6~eV for indentation rate and $\varepsilon$. These changes promoted the formation of the flake and we consider this case in more detail.

\begin{figure}[htb]
\centerline{\includegraphics[width=0.5\textwidth]{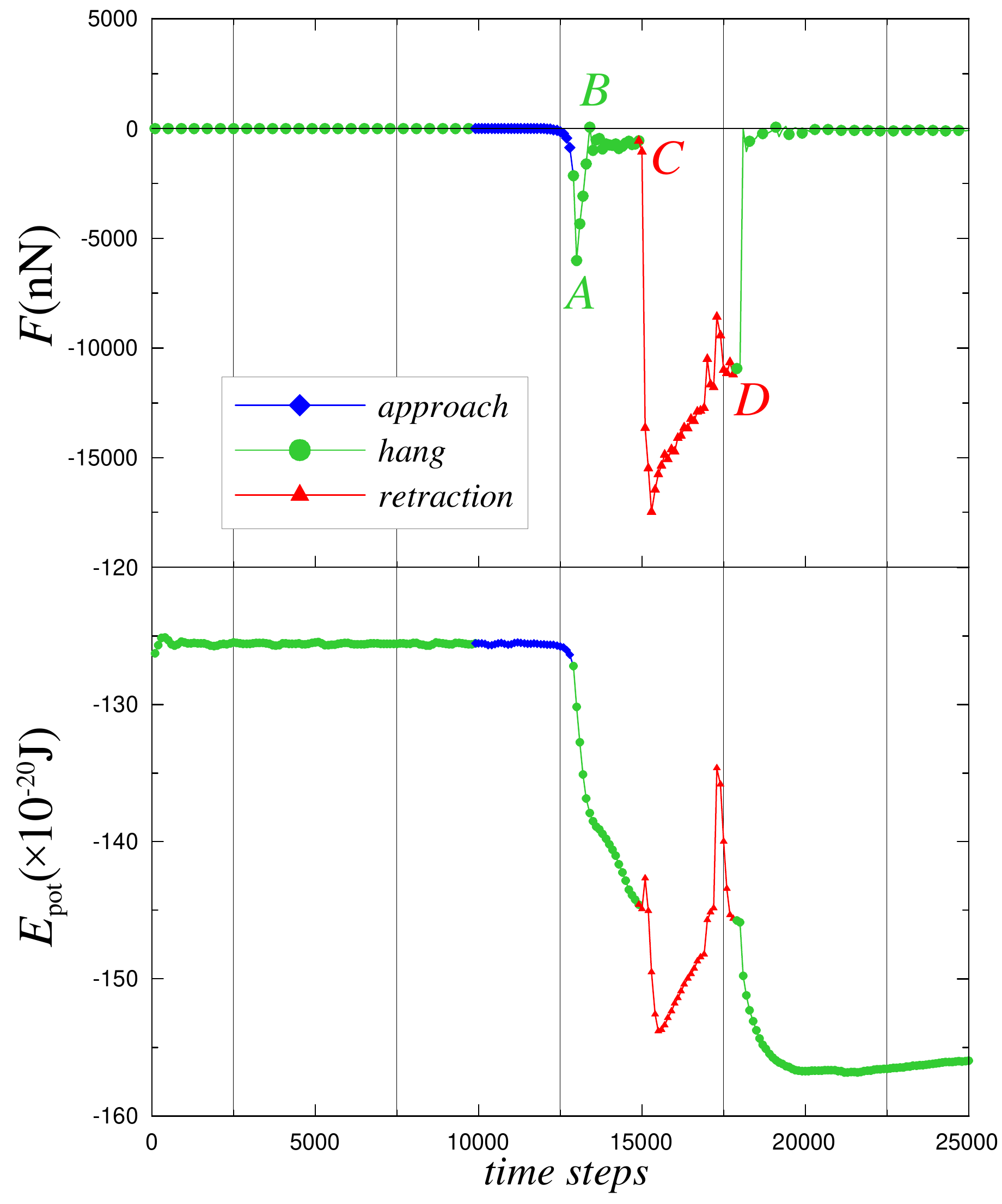}}
\caption{Time dependencies of the normal force acting on the tip and the potential energy of the system (per atom) for contact.}
\label{fig10}
\end{figure}

\begin{figure}[htb]
\centerline{\includegraphics[width=0.51\textwidth]{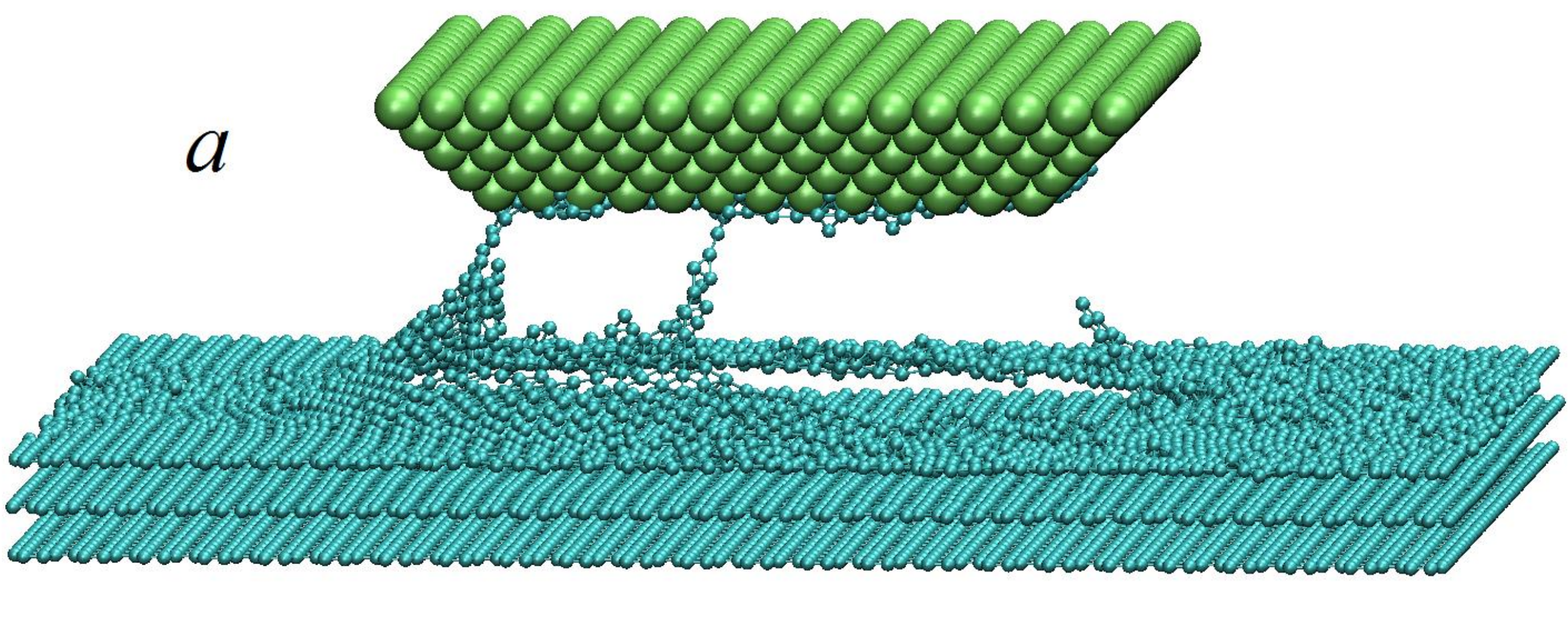}
\includegraphics[width=0.51\textwidth]{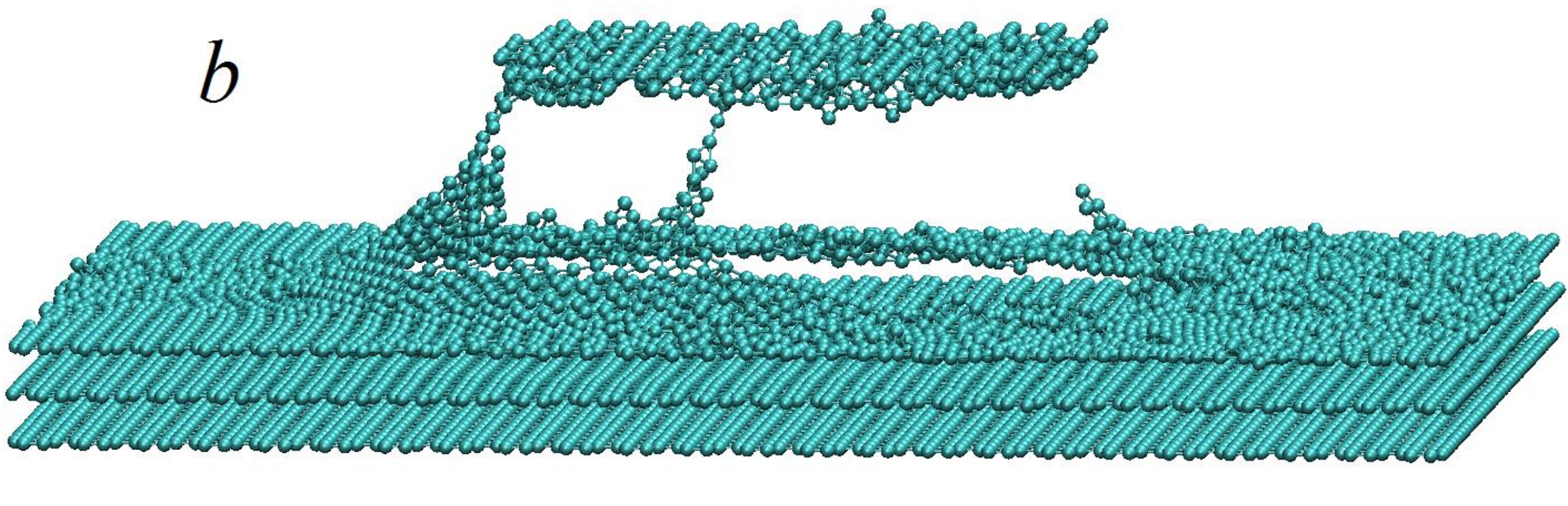}}
\caption{Configurations of the system with (\textit{a}) and without (\textit{b}) the tip at the end of the simulation of contact. The upper layer is teared and the flake is attached to the nanoasperity.}
\label{fig11}
\end{figure}

When the tip is pulled toward the surface, a gradual increase of the attraction between the two is observed, which is followed by a jump-to-contact (point \textit{A} in fig.~\ref{fig9} and fig.~\ref{fig10}). A sudden jump in force is represented by segment \textit{AB} as in the case of indentation. Retraction begins after point \textit{C} and causes a fast increase of the attraction, which diminishes till point \textit{D}. Note the absence of large scattering of data points in contrast to what was observed for indentation. This may be attributed to strong tip-substrate coupling and to the high rate which do not allow carbon atoms to collide with nanoasperity in between of its displacements. As animated video shows, the decrease in attraction is caused by rupture of interatomic bonds in the upper graphene layer. The breaking of the majority of bonds leads to the formation of a flake, signified by the considerable drop of the force after point \textit{D}. The final configurations of the system are shown in fig.~\ref{fig11}. During this simulation the temperature rises to about 1280~K, which is approximately 4.3 times larger than during indentation.

As can be seen from fig.~\ref{fig11}, the flake is very deformed and has almost completely lost its honeycomb lattice. Atoms in the upper graphene layer are also rearranged in a structure, which is distinct from the initial one. The reasons for this are a very high surface energy of the tip caused by a large value of $\varepsilon$ and the use of the bond order function from the first version of the Brenner potential. Mentioned facts impose evident constraints on our model towards complete reproduction of the superlubricity phenomenon, as it is assumed to base on the symmetry of the graphitic honeycomb lattice. It should be noted that it is unlikely to experimentally observe considered scenario of flake formation which occurs relatively far away from the graphene edges, because it requires very high indentation rates and surface energies of the tip (although magnitudes of the rate in the simulations may be influenced by strong coupling to the thermostat). More probable is the cleavage occuring at a grain boundary of a polycrystalline graphite sample, which is used in the experiments~\cite{Dien2004}.

\section{Conclusions}
Molecular dynamics simulations reported in the present study are capable of elucidating the general peculiarities of exfoliation of graphite in spite of a comparatively large number of approximations involved in the model such as consideration of the crystalline nanoasperity instead of amorphous one, the absolute rigidity of the tip and the bottom graphitic layer, the use of a simplified version of the Brenner potential and high indentation rates. The described computer experiments pave the way towards the completely realistic reproduction of the experiments pertaining to superlubricity. Obtained results may also provide some insights concerning the process of micromechanical cleavage of bulk graphite which is used for the production of graphene~\cite{Geim2007}.


\textbf{Acknowledgements}

We acknowledge the use of a Windows beowulf cluster with 12 dual-core processor nodes, which was provided by Informatics department at Sumy State University.

\end{document}